# The evolution of next-generation sequencing technologies

Olaitan Akintunde, Trichina Tucker, and Valerie J. Carabetta[*]

Department of Biomedical Sciences, Cooper Medical School of Rowan University, Camden NJ, 08103

*Correspondence: carabetta@rowan.edu. Tel: 856-956-2736

**Abstract**

The genetic information that dictates the structure and function of all life forms is encoded in the DNA. In 1953, Watson and Crick first presented the double helical structure of a DNA molecule. Their findings unearthed the desire to elucidate the exact composition and sequence of DNA molecules. Discoveries and the subsequent development and optimization of techniques that allowed for deciphering the DNA sequence has opened new doors in research, biotech, and healthcare. The application of high-throughput sequencing technologies in these industries has positively impacted and will continue to contribute to the betterment of humanity and the global economy. Improvements, such as the use of radioactive molecules for DNA sequencing to the use of florescent dyes and the implementation of polymerase chain reaction (PCR) for amplification, led to sequencing a few hundred base pairs in days, to automation, where sequencing of thousands of base pairs in hours became possible. Significant advances have been made, but there is still room for improvement. Here, we look at the history and the technology of the currently available high-through put sequencing platforms and the possible applications of such technologies to biomedical research and beyond.

1. A Brief History of DNA Sequencing

Deoxyribonucleic acid (DNA) is made up of four nitrogenous bases, which encodes information that provides the blueprint for the cell and directs physiology. Proper cell functioning requires proper DNA transcription and eventual translation into proteins. Any mutation or alteration in the DNA sequence in the genome could lead to possible cellular defects or disease. The early attempts to determine the identity and correct order of bases present in a DNA fragment, referred to as 1st-generation sequencing, involve the use of chemicals to cleave bases within a DNA molecule or the use of chain-terminating nucleotides, followed by the manual separation of fragments generated via electrophoresis [1,2]. Maxam and Gilbert sequencing, also called the partial chemical degradation method, utilizes chemicals that target specific, individual purines or pyrimidines to cleave the radiolabeled DNA backbone into fragments. The DNA is labelled with radioactive $^{32}$P at the 5' end. Next, specific chemicals are used to modify the nucleotides. For example, hydrazine removes the nitrogenous base from cytosine and thymine, but in the presence of high-salt concentrations, it preferentially cleaves cytosine. Formic acid is used to methylate adenine and guanine is methylated by dimethyl sulfate. Following base modification, piperidine is used to cleave the sugar-phosphate backbone, producing fragments that are analyzed using polyacrylamide gel electrophoresis [3].

Later, Sanger sequencing was introduced, which uses dideoxynucleotides (ddNTPs). ddNTPs are chemically modified dNTPs that lack hydroxyl groups at both the 2' and 3' positions and terminate extension once incorporated. This methodology requires a single-stranded DNA (ssDNA) template and a short oligonucleotide primer. Each ddNTP is added to the reaction at a fraction of the concentration of regular dNTPs, in the presence of the Klenow fragment for synthesis of the complementary strand. When a ddNTP is incorporated by the polymerase, chain termination occurs (Figure 1). Four parallel reactions are carried out yielding fragments of varying lengths, each ending with the specific ddNTP in use. This technique was originally performed with radiolabeled nucleotides, but the development of fluorescence labeling now allows for fluorometric detection [1,4]. The read length using this methodology is typically less than 1 kilobase (kb), which allows for low-throughput sequencing.

To sequence larger fragments or entire genomes, the DNA must be fragmented, cloned, individually sequenced, and then computationally assembled into genomes. Automated Sanger sequencing was first introduced by Applied Biosystems (ABI) in 1986, following the unveiling of the ABI Prism 310, a commercial DNA sequencing machine that employed automated capillary electrophoresis and an advanced imaging system. Four separate reactions, each with one of the four ddNTPs, were combined on a single gel and fluorescence was measured. A computer program was designed to interpret and compile the sequences, removing the need for manual sequence interpretation, reducing human error [5]. This automation allowed for early error detection and higher resolution [6]. Craig Venter and colleagues established the first, fully equipped sequencing facility, with six automated sequencers at the National Institutes of Health (NIH). In the 1990s, larger facilities were established, like the Institute for Genomic Research (TIGR), which had 30 sequencers and the Welcome Trust Sanger Institute. The availability of automated sequencing led to the discovery of 337 new and 48 homologous human genes via the expressed sequence tag (EST) method. ESTs are short DNA sequences (<500 bp) obtained from the 5' and 3' regions of cDNA. They are used as an efficient tool for the rapid identification of coding sequences in a genome by sequence matching, which reduces time and costs compared to other methods. This approach was widely used in the search for genes linked to human diseases, such as Huntington's disease. The National Center for Biotechnology Information (NCBI) has an EST database, a collection of annotated ESTs that serve as a gene discovery tool [7].

Later, next-generation sequencing (NGS), also referred to as high-throughput sequencing (HTS) or $2^{nd}$ generation sequencing, was developed. NGS technology started with the development of pyrosequencing [8] and was first commercially available in 2005 as the 454/Roche platform [9]. NGS utilizes a massive parallel sequencing approach, with the ability to process large DNA samples at a significantly reduced cost and time. One downside of Sanger sequencing is that it is relatively low throughput and costly. For example, the first human genome sequence cost approximately 1 billion dollars [10]. With NGS, millions or even billions of reads can be produced in a single run within a few hours or days, making it more efficient than Sanger sequencing [11,12]. The development of reversible dye terminator technology was a large contributing factor to the success of NGS technology [13]. Unlike sanger sequencing, which uses ddNTPs lacking the 3'-OH group to terminate extension irreversibly, this technology utilizes modified nucleotides attached to a reversible termination group. This group includes a fluorescent dye attached to a nucleotide via cleavable linker and can be photocleaved using ultraviolet irradiation after the nucleotide is incorporated and identified. This allows for the continuation of DNA extension by incorporation of the next nucleotide [14]. Photocleavage of the tags eliminates the need to use chemical reagents in the reaction, yielding a cleaner product without additional purification steps.

The human genome (~3 billion bp) sequencing using the 454 Genome Sequencer FLX was completed in 2 months, with a cost of about $100,000. In contrast, it took 15 years with Sanger sequencing, required multiple collaborations across the world, and was costly [15,16]. A shortcoming of NGS is the inability to read the entire length of the genome sequence, as millions of short fragments or reads are generated. Therefore, it takes a lot of computing power to piece these short reads together into one large sequence [12]. Recently, the development of new technologies for longer reads was developed, which can sequence double-stranded DNA one strand at a time and yield read lengths of > 10 kb. These technologies eliminate the need for amplification and make downstream genome assembly easier. While there will always remain some imperfections, notable improvements have been made and eventually will be a great option for disease diagnosis, by detecting structural variations linked to pathology [2]. In this review, we explore the popular NGS platforms currently being used. We discuss the diverse applications of NGS, including whole genome sequencing, and transcriptomics. We highlight the strengths and limitations of each approach and discuss some clinical applications.

## 2. Popular Next-generation Sequencing Technologies

DNA sample preparation, immobilization, and sequencing are the three critical steps of NGS technology. Sample preparation for sequencing usually involves attachment of specifically defined adaptor sequences to the ends of random DNA fragments. The resulting products of this DNA preparation is commonly referred to as the "sequencing library." The addition of adapters is necessary to secure the DNA fragments of the sequencing library to a solid platform and specify the site where the sequencing reaction is initiated. Some NGS platforms are largely based on pyrosequencing, like the Roche 454 series. Pyrosequencing is a technique where a DNA fragment is attached to a bead in an emulsion and following emulsion PCR, multiple copies of the same DNA fragment surround the bead. Luminescence is used to detect pyrophosphate ($PP_i$) release as a nucleotide is incorporated. ATP sulfurylase in the reaction mixture converts $PP_i$ into ATP, which is used by the enzyme luciferase to produce light. Advantages of this technique are that it uses natural nucleotides instead of modified ones for chain termination and allows for observation in real time. A modified version of this technique, known as ion torrent, measures the pH change for sequence determination [10,15].

Other NGS platforms, such as Illumina, require the use of a reversible dye terminator that allows for the addition of one nucleotide at a time during the DNA extension phase [1,16]. Illumina uses bridge amplification, in which adapters ligated on both ends of a fragment will bind oligonucleotides on a flow cell, which bend and form a loop or bridge structure. This loop serves as a primer for reverse strand synthesis (Figure 2). Multiple rounds of extension yield millions of fragments and the original sequence is determined by the reversible dye terminator method [13,17]. Another popular platform is SOLiD (Small Oligonucleotide Ligation and Detection system) sequencing, that sequences oligonucleotides with fixed lengths by ligation of fluorescent molecules. DNA ligases are used to add fluorescent oligonucleotides that anneal to DNA templates in a sequential fashion [15,18]. Multiple ligation cycles to a fluorescently labeled probe containing dinucleotides allows for sequence determination two nucleotides at a time [19]. In the next sections, we will describe four commonly used NGS platforms. For a recent, detailed review of the performance of the specific benchtop instruments developed by each company, see [20].

*2.1 454 GS FLX Titanium (Roche)*

The 454 GS FLX Titanium sequencer utilizes the pyrosequencing method, which was first introduced in 1996, by Nyren and Ronaghi at the Royal Institute of Technology, Stockholm [8]. Both Sanger's dideoxy and this pyrosequencing method are sequence-by-synthesis (SBS) techniques, as DNA polymerase is required to generate the desired output. Pyrosequencing is based on generating a complementary strand of ssDNA, while concurrently releasing a signal from the incorporated nucleotides. With pyrosequencing, one nucleotide at a time is passed over multiple copies of the template DNA to be identified, leading to nucleotide incorporation by polymerases via complementary base pairing. Following the synthesis of the longest possible complementary strand by the polymerase, base incorporation will end. One $PP_i$ molecule is released per nucleotide incorporated, followed by ATP sulfurylase-driven conversion of pyrophosphate to ATP, which emits light in the presence of luciferase. This works by a reaction of ATP with the substrate luciferin, to yield luciferyl-adenylate, which later reacts with oxygen to produce the light emitting oxyluciferin [21]. The light intensity is measured after base incorporation, followed by removal of unincorporated nucleotides, after which, the next nucleotide is introduced [8]. To protect the consumption of dATP by luciferase during the sequencing reaction, a modified nucleotide, deoxyadenosine-5'-($\alpha$-thio)-triphosphate (dATP$\alpha$S), which is unusable by luciferase, is used during base incorporation. The other three nucleotides are standard dNTPs.

Pyrosequencing technology evolved into parallel sequencing on a picotiter plate in 2005, first by 454 Life Sciences, which was later acquired by Roche Diagnostics (Table 1, [9]). This plate harbors about two million wells, each of which can hold one 28-μm diameter microbead. DNA from a sample is fragmented and each ssDNA fragment is attached to a single microbead via adapters. The surface of each bead contains oligonucleotides that bind to the adapter end, securing the DNA. Emulsion PCR (emPCR) of each bead produces a DNA library with millions of copies of that fragment. Clonal amplification occurs during emPCR, as each unique template molecule is physically separated from all others, with daughter molecules remaining bound to the microbeads. The plate is then incubated and washed with each dNTP, applied sequentially, in the presence of ATP sulfurylase and luciferase. $PP_i$ release is detected from light emission that is captured during nucleotide incorporation using a high-resolution, charge-coupled device (CCD) camera under the well [9,10]. The read length for 454/Roche is 400-500 bp. The average substitution error rate, excluding insertions-deletions (InDels), is in the range of $10^{-3}$ –$10^{-4}$. While this rate is higher than that of Sanger sequencing, it is among the lowest average substitution error rate of the newer sequencing technologies [16].

*2.2 Ion-torrent (Life Technologies)*

This platform detects and measures hydrogen ion release or pH change following nucleotide incorporation instead of fluorescence. Like 454 sequencing, beads covered with multiple copies of a DNA fragment produced via emPCR are added to a picotiter plate. Next, DNA polymerase and one nucleotide at a time are added, with nucleotide incorporation measured via pH change driven by proton release, in place of $PP_i$ release (Figure 3). This technology is driven by complementary metal-oxide-semiconductor (CMOS) technology used in the microprocessor chips production. This is a low-cost semiconductor that allows for non-optical DNA sequencing by using sensors to detect ions produced during polymerase-driven sequencing reactions, which enables rapid sequencing during the actual detection phase. The read lengths of the ion-torrent is 200-600 bp, depending on the specific instrument available (Table 1, [1,22]).

*2.3 Illumina sequencing*

The Illumina sequencing process involves adaptor-ligated DNA fragments immobilized on a glass surface that are subjected to clonal amplification [23]. The reversible dye terminator technique is the foundation for the Illumina Genome Analyzer (GA). This platform utilizes a sequencing-by-synthesis approach. Like the 454, the protocol requires conversion of DNA into an adapter-ligated library and immobilization onto a surface for sequencing [16,23,24]. Sodium hydroxide is used in the denaturation of the dsDNA library. The resulting ssDNA molecules, at a low concentration, are pushed through the channels of a flow cell with fixed oligonucleotides on a glass surface that will anneal to the two different adapters attached to the 5´ and 3´ ends, based on complementarity. This arrangement allows for bridge amplification, in which the synthesis of the reverse strand starts from the annealed portion and the newly synthesized strands are covalently attached to the flow cell (Figure 2, [25]). The tethered DNA molecule then bends and binds to another oligonucleotide complementary to the second adapter on the open end of the strand, generating a second covalently attached reverse strand cell (Figure 2). Several bridge amplifications create clusters of both forward and reverse strands, consisting of thousands of copies of the original sequence, closely packed on the flow cell [10,16,23,24]. After denaturation, each cluster will have only single stranded, identically oriented copies of the original DNA molecule. Visualization of the clusters produced by bridge amplification occurs by detection of fluorescent reversible-terminator nucleotides. In this approach, fluorescently labeled 3´-O-azidomethyl-dNTPs are used to pause the polymerization reaction, enabling the removal of unincorporated bases and fluorescent imaging to determine the identity of the added nucleotide [26]. Following scanning of the flow cell with a CCD camera, the fluorescent moiety and the 3´ block are removed and the process is repeated. Here, the read lengths are shorter than for Sanger sequencing, but it has increased sequencing speed and output. For examples, using this technology millions of reads are obtained within hours. It is also more cost efficient and eliminates the need for gel electrophoresis (Table 1). Illumina is among the most widely used and arguably the most successful sequencing technology [1].

*2.4 SOLiD (Applied Biosystems)*

Pioneered by Harvard Medical School and the Howard Hughes Medical Institute in 2005 and unlike other methods where the DNA extension reaction is executed by polymerases, SOLiD utilizes ligases (Table 1). The ligases attach fluorescently labeled primers in a sequence-specific manner to the template strand [18,19]. Once the initial universal sequencing primer has been annealed to a single-stranded template of DNA molecules in a library, the primer is extended by introducing a mixture of octamer probes. Each probe contains four unique $5^!$ fluorescent labels, with one of the 16 possible dinucleotide combinations (e.g., AT, AG, AC, etc.) at the $3^!$ end. The octamers compete to be ligated to the sequencing primer based on complementarity to the template sequence. The last two complementary nucleotides at the $3^!$ end of the probe are read. On the $5^!$ end of the probe, three bases and the dye present are removed, exposing a free $5^!$ phosphate for continued extension. After 10 cycles of ligations or the desired length is reached, the extended product denatured and a new sequencing primer, shifted by one nucleotide is annealed to the template to repeat the process. This same process is repeated three more times with three additional sequencing primers, each shifted by one nucleotide compared to the previous. Performing five rounds of ligations with five different, shifted primers allows for dual measurement of each base in the sequence and therefore, increases sequencing accuracy [16,18].

3. **Single molecule sequencing**

Single molecule sequencing (SMS), also known as third-generation sequencing, is executed without fragment amplification, so both DNA strands can be sequenced, providing more information and increased accuracy [27]. Advantages of SMS are that it reduces errors that occur during DNA amplification or the library preparation step and is effective to analyze low DNA concentrations. In addition, this technique can identify non-standard nucleotides, like modifications such as methylation, that are omitted during the amplification step of the other platforms [16]. Another benefit is the generation of longer reads compared to other NGS platforms (Table 1). Longer reads reduce the computing requirements of fragment assembly during applications such as whole-genome sequencing. A final advantage of this technology is that it can also detect epigenetic markers without the need for chromatin immunopurification. The downside is that SMS techniques have a higher error rate in comparison to other NGS technologies. Next, we will discuss the two common SMS platforms that are readily in use.

*3.1 Single-molecule real-time (SMRT) sequencing (Pacific Biosciences)*

In SMRT sequencing, nucleotide detection occurs in tiny chambers called zero-mode waveguides (ZMWs). DNA polymerase is fixed to the bottom of each ZMW, to which target DNA and nucleotides tagged with different colored fluorophores are added [28]. Single DNA strand extension can be monitored in real time, as only incorporated nucleotides emit fluorescence. This eliminates interference from other unincorporated labelled dNTPs in solution. The ZMW is an optical waveguide that guides laser energy through pores with diameter narrower than the wavelength. Therefore, the electromagnetic energy from the excitation beam decays as it penetrates the nanoscale aperture. This allows a small detection volume, on the order of 100 zeptoliters, that helps reduce background noise. In other words, a single incorporation event is easily captured against a background of other nucleotides [10,11,28]. For this technique, read lengths are significantly increased, and typically will be 14–60 kb (Table 1).

*3.2 Nanopore DNA sequencing (Oxford)*

For this SMS platform, ssDNA is electrophoresed by a molecular motor protein, such as DNA helicase. The DNA helicase binds to ssDNA and pushes it through a biological nanopore embedded in a synthetic membrane, across which a voltage is applied (Figure 4). As the ssDNA passes through the nanopore, individual bases disrupt current flow in a unique manner, allowing the sequence of the molecule to be inferred by monitoring the changes in current. In other words, the nucleotide movement through the pores impedes the flow of ions, reducing current for a time proportional to the DNA length and composition of nucleotides. This allows for the sequencing of both strands of a DNA molecule [10,29,30].

An advantage of the Nanopore is the long read lengths achievable, with different modes capable of detecting 10-300 kb (Table 1).

## 4. Applications of NGS Technologies

To better understand the underlying pathophysiology of a disease, there may be a need to study mutations in genes, but also the expressed RNAs and proteins to provide clues to its management and treatment [23]. Whole-genome sequencing allows for the detection of mutations or genetic variants that may be responsible for a particular disease. Transcriptomic analysis will aid in understanding exactly which genes are expressed at different times and can also be used to monitor disease progression. Advances in sequencing technology did not just reduce the cost and streamline the process; they have also paved the way for new applications to elucidate molecular mechanisms of genomic structure and cellular functions [15,31]. Next, we discuss some traditional and newer applications of NGS technologies. We highlight the methodology, clinical or research applications, and discuss any noted limitations of each approach.

*4.1 DNA sequencing techniques*

Detecting DNA alterations that affect human health is now possible because of NGS technologies. Some techniques that can be applied to the study of disease are DNA sequencing (DNA-seq [32]), single cell DNA-seq (scDNA-seq), whole exome sequencing (WES, [33]), and targeted sequencing (TS, [34]). Sequencing of the entire genome provides global information about exons and introns, which can reveal the regulatory components of genes, such as promoters, enhancers, and intronic regulators, and structural variants, like copy number variants, inversions, and translocations [35,36]. The read depth, meaning the number of times each nucleotide is detected in a sequencing reaction, is crucial for sequencing accuracy. To observe small changes, such as single nucleotide substitutions (SNPs) or point mutations linked to diseases, a high read depth and sequencing of multiple affected individuals are required. However, if there are large structural changes relative to a reference genome, information is achievable even with a low read depth. In most cases, increasing read depth can be expensive and cheaper methods, such as customized hybridization chips, may be preferred. One issue is that most sequencing runs yield a limited number of fragments reads. For larger genomes, a more cost-effective approach is WES, in which only the DNA sequence that is transcribed into mRNA (exons) are analyzed. WES is selective and enriches for sections of genomic DNA captured using commercially available capture arrays. These arrays harbor unique baits to capture target exons. Frequently, streptavidin or magnetic beads coated with oligonucleotides that bind and capture exome sequences are used. A benefit of WES for analyzing the human genome, is that the price can be significantly reduced since only about 1% of the ~20,000 human genes are coding sequences. The downside is that the bait oligos can sometimes miss unknown coding regions or regions with key mutations that could be important for an underlying disease state [31,36].

TS is another popular technique, which allows for a gene or region of a genome to be assessed for variations that may be linked to a phenotype or specific disease. It is faster, more accurate, less expensive and requires less computing power than whole-genome sequencing. Sequencing a small number of target genes typically requires PCR amplification, followed by NGS. For many diseases, multiple genetic panels are predesigned or customized to isolate the desired genomic regions by hybridization probes. The high read depth of TS, considered ultra-deep sequencing, makes it possible to identify rare mutations present in only a small subpopulation of cells, which is highly advantageous for the study of malignancies. TS has become a mainstay in cancer research and treatment [37-39]. Identification of genetic variations linked to a disease or response to drug treatment is essential for optimizing drug design or individual pharmacotherapy [37,40]. Another popular technique used for the study tumor development and progression is scDNA-seq [41,42]. This technique involves isolation of single cells from a population utilizing either low throughput methods, like serial dilution, laser capture microdissection or fluorescence-activated cell sorting (FACS), or high throughput methods, such as microfluidic chips or combinatorial indexing [43]. The cells are then lysed to recover genetic material. A sequencing library is prepared by fragmenting the DNA and ligating adapters for NGS. While DNA-seq provides the sequence of the average genomic sequence of a population of cells, single cell sequencing shines light on the genome of a single cell within a population. scDNA-seq is becoming increasingly popular and is well suited for the study of mutational processes, mosaicism, lineage tracing, germ line mutations, and cancer progression [44].

As with any technology, DNA sequencing has its limitations. One common problem encountered is the underrepresentation of high GC- or AT-rich regions. There is often an inherent difficulty in differentiating sequencing errors from actual variants within a population, especially if they are rare. Mapping errors frequently occur with short read alignments around repetitive chromosomal regions. In addition, errors occur when arranging short sequencing reads into larger assemblages and there are problems in identifying large structural variants, like inversions. Despite these limitations, DNA sequencing technology has revolutionized our understanding of cellular physiology in health and disease.

*4.2 RNA sequencing techniques*

Early attempts to understand and quantify transcript levels on the global scale relied on microarray technology. Microarray technology is an approach where a desired sample is bound to a surface or chip containing thousands to millions of immobilized nucleic acid fragments. mRNA is recovered from different samples or individuals and reverse transcribed to cDNA, which are then exposed to the microarray chip to determine which genes are expressed. One downslide to microarrays is that they require a sequenced reference genome and transcriptome for its design and interpretation. NGS techniques, such as RNA-seq, can reveal the expression profile of organisms with un-sequenced genomes [45]. RNA-seq is more sensitive, provides better accuracy, and is not influenced by chip sequence biases [15]. It relies on high-throughput sequencing of cDNA fragments produced from RNA or RNA fragments, which results in an accurate mapping of transcripts to unique regions of the genome and eliminates most of the background noise. As transcripts can be precisely quantified, it is possible to analyze transcript isoforms within a 5000- fold dynamic range [15,46]. RNA-seq has led to the discovery of novel genes and RNAs that where not detected by microarray technology, such as novel gene fusions [47]. It can also reveal possible differences between exomes and the transcriptome caused by RNA-editing [48].

For this technique, mRNA is reversed transcribed to cDNA, following careful and rigorous isolation techniques that yield quality mRNA and often deplete rRNA. As rRNA represents the majority of the RNA species in a cell, depletion can lead to the identification and quantification of lower abundance RNAs. Different RNA preparation methodologies have emerged over time, but all are based on purifying desired mRNA, non-coding RNAs, or miRNAs from other RNA species and contaminants. A typical workflow involves the selection of polyadenylated mRNA, followed by rRNA depletion, and reverse transcription. However, this approach is biased for the 3' end, works better with high quality mRNA, and yields poor data recovery from non-coding portions of RNA [49,50]. RNA-seq with prokaryotic RNAs is possible, even though they lack poly-A tails and cannot be enriched or selected with poly T primers. This requires the use of random oligonucleotides, often times hexamers consisting of six random nucleotides [51]. Data analysis and interpretation occur via freely available bioinformatic platforms. Sequencing reads are aligned to a reference genome and assembled into transcripts. Different software packages have been specifically designed for RNA-seq analyses [52-54].

RNA-seq has been used for many applications in medicine, including the study of disease progression overtime, such as heart failure [45]. RNA-seq is commonly used to study differential gene expression [55], which can reveal functions of new genes or identify new cancer biomarkers [54,56,57]. An extension of RNA-seq is microRNA (miRNA)-seq. miRNAs are a class of RNAs, typically ~22 nucleotides in length, that regulate multiple physiological and pathological processes. Their main function is to regulate translation and stability of their target mRNAs. The study of miRNA expression patterns in cells and tissues has been used to study development and disease pathology, especially in cancer cells [58,59].

As with DNA sequencing, single cell RNA-seq (scRNA-seq) is also being increasingly performed to study how gene expression varies among individual cells within a population [60]. Traditional RNA-seq measures gene expression across a population and does not address population heterogeneity. This has been especially powerful in the study of the underlying pathogenesis of cancer cells. Tumor cell heterogeneity is believed to be a significant cause of treatment failure, so this understanding is necessary to optimize treatment regimens [61].

4.3 *NGS for the study of epigenetics*

NGS technology has been developed to study epigenetics, which are reversible changes to nucleosomes or nucleic acids that produce phenotypic variation without corresponding alterations to the genome. In general, eukaryotic DNA is organized into actively transcribed regions called euchromatin and transcriptionally silent regions called heterochromatin. These loosely or tightly packed regions depend on the combination of modifications present on the N-terminal tails of histone proteins, which make up the nucleosomes, and other silencing factors [23,62]. A technique to identify regions of euchromatin is Assay for Transposase-Accessible Chromatin (ATAC)-seq [63], which utilizes a hyperactive Tn5 transposase mutant to introduce sequencing adapters into the actively transcribed, open regions of the native chromatin. The tagged sequences are purified, amplified, and sequenced using NGS technologies. Sequencing reads are used to identify regions with increased accessibility and are used to map binding sites of histones. ATAC-seq can determine the location and interactions between transcription factors, nucleosomes, and open chromatin, and shed light onto how chromatin compaction is regulated [63].

Another technique to study epigenetic modifications is whole-genome bisulfite sequencing (WGBS). WGBS detects the presence of 5-methylcytosine, which is an abundant epigenetic DNA modification that is typically associated with gene silencing [64]. For this technique, extracted genomic DNA is treated with sodium bisulfite to convert unmethylated cytosines into uracil, while methylated cytosine will remain unchanged. During the PCR step, the uracil bases are changed to thymine, which means that the sequencing results will be composed of mostly thymine, adenine, and guanine. The presence of cytosine indicates the methylation sites [65]. WGBS has been successfully used to study stem cell differentiation, development, disease diagnosis and forensic science [66-68].

*4.4 NGS for the study of DNA-binding proteins*

Chromatin immunoprecipitation sequencing (ChIP-seq) is a widely used technique for mapping protein-DNA interactions. For this methodology, DNA-binding proteins are crosslinked to their target DNA binding sites often using chemical crosslinkers, like formaldehyde, and the DNA is typically sheared by sonication or enzymatic treatment with micrococcal nuclease (MNase). The immunoprecipitation step requires a specific antibody directed against the DNA-binding protein of interest. The use of a high-quality, efficient antibody or ChIP-certified antibodies is necessary to reduce the signal to noise ratio to ensure quality data. The crosslink is reversed, often through extensive heat treatment, proteins and RNA degraded by treatment with appropriate enzymes, and target DNA isolated. A sequencing library with appropriate adaptors is then produced [65]. Peak detection software maps sequencing reads to a reference genome and detects regions with enriched frequency (different software tools reviewed in [69]). Applications of ChIP-seq include mapping transcription factor binding sites, studying chromatin structure, and epigenetics [70]. In fact, ChIP-seq is often used for annotating the state of chromatin under specific conditions and to study histone modifications [71]. By performing the immunopurification with antibodies specific for different histone modifications, more information on which chromosomal regions are enriched with specific modified species is available. Modified histones are promising as biomarkers for various diseases and ChIP-seq is an effective technique for their study [72].

*4.5 NGS for the analysis of chromosome structure and positioning*

Tyramide signal amplification (TSA)-seq has been described as a cytological ruler that measures intracellular distance between the chromosome and intranuclear structures, enabling the spatial understanding of the three-dimensional genome organization. It can reveal distances between genetic loci and nuclear compartments and provide information on how the chromosome spans or travels across those different compartments. The proximity of a gene to the center of the nucleus can influence its activity and may change as the cell grows. This method involves the use of primary antibodies directed against specific proteins of interest combined with secondary antibodies that are coupled with the enzyme horseradish-peroxidase (HRP). Following cross-linking with formaldehyde, cells are labeled with biotin-tyramide and HRP catalyzes the formation of biotin-tyramide free radicals, which creates a concentration gradient with a radius of 1 $\mu$m. TSA labels both proteins and DNA. DNA fragments in close proximity will be labeled and isolated by immunopurification of biotin with streptavidin beads. The signal intensity increases as the distance between the protein of interest and the chromosome decreases [73].

Another technique used to study long-range DNA-DNA interactions and chromosome architecture is chromosome conformation capture (3C) sequencing (3C-seq). This technique is used to identify the functional interactions among unique chromosomal segments [74,75]. In other words, 3C-seq, also called Hi-C, allows for the study of chromosomal loci that may interact when a genome is folded in three-dimensional space versus when it is linear. For this methodology, chromatin is cross-linked with formaldehyde, which creates covalent bonds between segments of DNA that are bridged together by proteins in 3D space. The DNA is digested with restriction enzymes, re-ligated and crosslinks reversed. This creates 3C templates which are used for NGS [76]. TSA-seq and 3C-seq have led to an unparalleled understanding of the 3D structure and organization of genome in living cells.

*4.6 NGS applications for microbial population studies*

Transposons are mobile genetic elements that can be excised and insert in different loci throughout a genome. The insertion of a transposon will most likely disrupt gene function and thus transposons became a powerful tool to study gene functions. Transposon sequencing (Tn-seq) is a technique which requires the generation and use of a transposon library in a microorganism. Transposons can only be inserted into genes that are non-essential for viability, but the fitness of the resulting mutant strains under specific growth conditions may vary greatly. Underrepresentation or overrepresentation of specific mutants in a population provides clues about their necessity in each specific condition. Mutant strains with a growth deficiency will be underrepresented in a population and contain an insertion in a gene that is conditionally essential. Mutants that are overrepresented in a population will represent insertions in genes that when deleted, confer a selective growth advantage. To identify the disrupted gene and determine the position and frequency of insertion in the population, the downstream genomic region from the 3' end of the insertion site is amplified. As with other techniques, adaptors are ligated to create a library, which is then sequenced using NGS technologies. Typically, sequencing primers designed internal to the 3' end of the transposon are used [77,78]. Tn-seq is one of the few techniques that can directly link genotype to phenotype in microbiology.

Transposon-directed insertion site sequencing (TraDIS) is an extension of Tn-seq, whereby amplification and sequencing of transposon rich fragment libraries are used to determine the rate of insertion. TraDIS uses random, mechanical shearing of the DNA by sonication, which improves mapping accuracy due to longer sequencing reads. This is different from Tn-seq, which uses restriction digests to produce uniform fragments during library preparation [79]. In the last decade, there have been even more advances in Tn-seq technologies, with novel combinations of techniques emerging. These include motility TIS, density-TraDISort, TraDISort and dTn-seq [80]. The first three methodologies involve separating cells based on phenotypes other than growth, like motility, density (capsule production) and FACS, respectively. Droplet Tn-seq (dTn-seq) utilizes microfluidics to sort single cells, in which individual mutant cells are encased in an oil droplet. The cells will grow in their own droplet, without the influence of other mutants in the population. As dTn-seq is compatible with FACS or microscopy, droplets can be re-encapsulated with other droplets containing different cells to allow for the signals to diffuse between the two. Sometimes, the true fitness of a mutant cannot be ascertained when it is in the presence of other mutants, which is when dTn-seq may be desired. In addition, the effects of secreted products or signals can be evaluated [81]. Another useful extension of Tn-seq is the TraDIS-Xpress technique. A limitation of traditional Tn-seq is that insertion of transposons in essential genes is lethal, so the study of their function is not possible. TraDIS-Xpress uses a Tn5 transposon where an inducible promoter is outward facing. Gene disruptions of non-essential genes will still occur, but insertion of the transposon upstream of an essential gene will lead to overexpression. This means that all genes can be analyzed using this technique [82].

## 5. Challenges and limitations of NGS technologies

As with any new methodology, there are great benefits, but there will always be limitations. Each sequencing technology has its own shortcomings and the capacity to detect variants differs [83]. One such limitation, especially in the DNA of higher eukaryotic organisms, is the high error rate when sequencing homopolymers, which are stretches of DNA containing multiple repeats of the same nucleotide. Homopolymers, also called mononucleotide microsatellites, are more difficult to sequence because of the absence of a blocking moiety in the nucleotides used, since the repeating nucleotide will completely incorporate during a single cycle, which leads to uncertainty in base calling. The sequencing of homopolymers is a challenge when using every NGS platform, especially for those which utilize sequencing-by-synthesis, such as the Ion Torrent and 454 [84]. Therefore, homopolymers may end up significantly shortened or expanded in the final genome assemblage. Homopolymers are not the only DNA sequence that is challenging. The accuracy of GC-rich regions remains a challenge for all sequencing platforms, including traditional Sanger sequencing. Furthermore, the short-read lengths of most platforms undermines the ability to correctly sequence large repeat regions and other large-scale variations that may be present in a genome [10]. Insertions, deletions, and larger structural variants are typically difficult to identify. Assembling of a *de novo* genome from short read assemblages remains a challenge and often leads to incomplete assembled sequences with gaps. Increasing read length and accuracy will significantly improve the coverage and accuracy of *de novo* genomes, which also will enable precise variant mapping between individuals [15].

As the number of platforms and applications increases, the need for standardization between researchers also increases. The National Institute of Standards and Technology (NIST) has been working with other agencies towards regulating NGS protocols by developing standards and metrics that are widely acceptable [85]. One area of focus is on the analysis human genomic DNA from multiple cell lines, which will serve as benchmarks and will be characterized for uniformity, stability, and quality using the available sequencing platforms and protocols. This will yield high confidence datasets that are reproducible and can be used by both clinical and research laboratories to better evaluate sequencing and bioinformatics techniques. Similarly, Brownstein *et al.* proposed the CLARITY challenge, to help compare the efficiency of various sequencing technologies. The CLARITY challenge was designed to standardize methods for diagnosing genetic disease using genomic sequencing data. In general, the participating researchers used similar workflows and downstream bioinformatic analyses. The study also identified important needs, like quality publicly available reference genomes and the importance of well documented clinical reports, including reference accession numbers and other bioinformatic statistics. The hope is to be able to set standards for diagnosing genetic diseases using both clinical case history and appropriately processed and analyzed genomic data [86].

6. Conclusions

Since the 1990s, considerable efforts to improve whole genome sequencing has led to significant advancements in sequencing capacity and output (Tables 1 and 2, [87]). Important advancements include:

1. the development of reversible dye terminators and mutant T7 DNA polymerase that allows for easy incorporation of such terminators [88], 2. amplification of DNA by PCR, which reduced the need for large quantities of template DNA and enabled small sample processing [89], 3. DNA fragment isolation and purification using magnetic beads enabled higher quality and purer starting material [90], 4. double-stranded DNA sequencing technologies, which allowed for cloned fragment and paired-end sequencing, increasing the read depth, 5. capillary electrophoresis, which eliminated manual gel electrophoresis and promoted fluorescence interpretation [91], and 6. automation of techniques, which led to a reduction in errors and improved efficiency. Like the NGS technologies, great strides have been made towards developing and improving software platforms to identify variants, quantify, and assemble large amounts of sequencing data. Software such as Phred eliminated the need for the manual editing of reads, improved read quality and helped analyze sequences with repeats [87]. Now, there are many freely available to software programs to analyze all the different types of data sets that are generated (reviewed in [92,93]).

With these tools, a small research group can generate large amounts of sequence data faster and cheaper than with conventional Sanger sequencing, with costs reduced to 0.1-4% and time shortened by a factor of 100–1,000 based on daily throughput. One challenge facing smaller research groups is the cost of acquiring these instruments. Recently, some smaller-scale, more cost-effective versions of the larger instruments have become available, like the Illumina GA IIe or 454/Roche GS Junior, but they have a lower sequencing capacity. However, the overall costs can still be rather high, as the cost of sequencing per base is increased compared to the larger instruments and still requires knowledge and familiarity with protocols. Sequencing technology choice depends on the experimental requirements and sometimes it may be necessary to combine multiple platforms. This scenario may create competition, allowing more companies to provide sequencing services as needed or required, which will ultimately save individual laboratories the time and expertise required for sequence generation and data analysis [16].

NGS technologies will continue to play a leading role in identifying new disease biomarkers and make importation contributions to our fundamental understanding of how genetic abnormalities contribute to different phenotypes [26]. Continuous efforts towards the improvement of these NGS platforms will ensure better accuracy and reproducibility, will reduce error rates, and the overall cost of sequencing [2]. As the number of potential applications and uses for NGS expands, these technological improvements will be necessary to keep up with the increased demands for faster turnaround times, higher accuracy, and a reasonable cost.

**Acknowledgements:** This work was supported by grant GM138303, awarded to V.J.C.

**References**

1. Heather JM, Chain B (2016) The sequence of sequencers: The history of sequencing DNA. Genomics 107 (1):1-8. doi:https://doi.org/10.1016/j.ygeno.2015.11.003

2. Hu T, Chitnis N, Monos D, Dinh A (2021) Next-generation sequencing technologies: An overview. Hum Immunol 82 (11):801-811. doi:10.1016/j.humimm.2021.02.012

3. Gilbert W, Maxam A (1973) The nucleotide sequence of the *lac* operator. Proc Natl Acad Sci U S A 70 (12):3581-3584. doi:10.1073/pnas.70.12.3581

4. Sanger F, Nicklen S, Coulson AR (1977) DNA sequencing with chain-terminating inhibitors. Proceedings of the National Academy of Sciences 74 (12):5463-5467. doi:doi:10.1073/pnas.74.12.5463

5. Smith LM, Sanders JZ, Kaiser RJ, Hughes P, Dodd C, Connell CR, Heiner C, Kent SB, Hood LE (1986) Fluorescence detection in automated DNA sequence analysis. Nature 321 (6071):674-679. doi:10.1038/321674a0

6. Hutchison CA, 3rd (2007) DNA sequencing: bench to bedside and beyond. Nucleic Acids Res 35 (18):6227-6237. doi:10.1093/nar/gkm688

7. Adams MD, Kelley JM, Gocayne JD, Dubnick M, Polymeropoulos MH, Xiao H, Merril CR, Wu A, Olde B, Moreno RF, et al. (1991) Complementary DNA sequencing: expressed sequence tags and human genome project. Science 252 (5013):1651-1656. doi:10.1126/science.2047873

8. Ronaghi M, Karamohamed S, Pettersson B, Uhlén M, Nyrén P (1996) Real-time DNA sequencing using detection of pyrophosphate release. Anal Biochem 242 (1):84-89. doi:10.1006/abio.1996.0432

9. Margulies M, Egholm M, Altman WE, Attiya S, Bader JS, Bemben LA, Berka J, Braverman MS, Chen Y-J, Chen Z, Dewell SB, Du L, Fierro JM, Gomes XV, Godwin BC, He W, Helgesen S, Ho CH, Irzyk GP, Jando SC, Alenquer MLI, Jarvie TP, Jirage KB, Kim J-B, Knight JR, Lanza JR, Leamon JH, Lefkowitz SM, Lei M, Li J, Lohman KL, Lu H, Makhijani VB, McDade KE, McKenna MP, Myers EW, Nickerson E, Nobile JR, Plant R, Puc BP, Ronan MT, Roth GT, Sarkis GJ, Simons JF, Simpson JW, Srinivasan M, Tartaro KR, Tomasz A, Vogt KA, Volkmer GA, Wang SH, Wang Y, Weiner MP, Yu P, Begley RF, Rothberg JM (2005) Genome sequencing in microfabricated high-density picolitre reactors. Nature 437 (7057):376-380. doi:10.1038/nature03959

10. Reuter Jason A, Spacek DV, Snyder Michael P (2015) High-throughput sequencing technologies. Molecular Cell 58 (4):586-597. doi:https://doi.org/10.1016/j.molcel.2015.05.004

11. Kchouk M, Gibrat J-F, Elloumi M (2017) Generations of sequencing technologies: from first to next generation. Biology and medicine 9:1-8

12. Ruparel H, Bi L, Li Z, Bai X, Kim DH, Turro NJ, Ju J (2005) Design and synthesis of a 3-*O*-allyl photocleavable fluorescent nucleotide as a reversible terminator for DNA sequencing by synthesis. Proceedings of the National Academy of Sciences 102 (17):5932-5937. doi:doi:10.1073/pnas.0501962102

13. Li Z, Bai X, Ruparel H, Kim S, Turro NJ, Ju J (2003) A photocleavable fluorescent nucleotide for DNA sequencing and analysis. Proceedings of the National Academy of Sciences 100 (2):414-419. doi:doi:10.1073/pnas.242729199

14. Turcatti G, Romieu A, Fedurco M, Tairi AP (2008) A new class of cleavable fluorescent nucleotides: synthesis and optimization as reversible terminators for DNA sequencing by synthesis. Nucleic Acids Res 36 (4):e25. doi:10.1093/nar/gkn021

15. Soon WW, Hariharan M, Snyder MP (2013) High-throughput sequencing for biology and medicine. Molecular Systems Biology 9 (1):640. doi:https://doi.org/10.1038/msb.2012.61

16. Kircher M, Kelso J (2010) High-throughput DNA sequencing – concepts and limitations. BioEssays 32 (6):524-536. doi:https://doi.org/10.1002/bies.200900181

17. Chen F, Dong M, Ge M, Zhu L, Ren L, Liu G, Mu R (2013) The history and advances of reversible terminators used in new generations of sequencing technology. Genomics Proteomics Bioinformatics 11 (1):34-40. doi:10.1016/j.gpb.2013.01.003

18. Shendure J, Porreca GJ, Reppas NB, Lin X, McCutcheon JP, Rosenbaum AM, Wang MD, Zhang K, Mitra RD, Church GM (2005) Accurate multiplex polony sequencing of an evolved bacterial genome. Science 309 (5741):1728-1732. doi:doi:10.1126/science.1117389

19. Voelkerding KV, Dames SA, Durtschi JD (2009) Next-generation sequencing: from basic research to diagnostics. Clin Chem 55 (4):641-658. doi:10.1373/clinchem.2008.112789

20. Pervez MT, Hasnain MJU, Abbas SH, Moustafa MF, Aslam N, Shah SSM (2022) A comprehensive review of performance of next-generation sequencing platforms. Biomed Res Int 2022:3457806. doi:10.1155/2022/3457806

21. Morciano G, Sarti AC, Marchi S, Missiroli S, Falzoni S, Raffaghello L, Pistoia V, Giorgi C, Di Virgilio F, Pinton P (2017) Use of luciferase probes to measure ATP in living cells and animals. Nature Protocols 12 (8):1542-1562. doi:10.1038/nprot.2017.052

22. Rothberg JM, Hinz W, Rearick TM, Schultz J, Mileski W, Davey M, Leamon JH, Johnson K, Milgrew MJ, Edwards M, Hoon J, Simons JF, Marran D, Myers JW, Davidson JF, Branting A, Nobile JR, Puc BP, Light D, Clark TA, Huber M, Branciforte JT, Stoner IB, Cawley SE, Lyons M, Fu Y, Homer N, Sedova M, Miao X,

Reed B, Sabina J, Feierstein E, Schorn M, Alanjary M, Dimalanta E, Dressman D, Kasinskas R, Sokolsky T, Fidanza JA, Namsaraev E, McKernan KJ, Williams A, Roth GT, Bustillo J (2011) An integrated semiconductor device enabling non-optical genome sequencing. Nature 475 (7356):348-352. doi:10.1038/nature10242

23. Bentley DR, Balasubramanian S, Swerdlow HP, Smith GP, Milton J, Brown CG, Hall KP, Evers DJ, Barnes CL, Bignell HR, Boutell JM, Bryant J, Carter RJ, Keira Cheetham R, Cox AJ, Ellis DJ, Flatbush MR, Gormley NA, Humphray SJ, Irving LJ, Karbelashvili MS, Kirk SM, Li H, Liu X, Maisinger KS, Murray LJ, Obradovic B, Ost T, Parkinson ML, Pratt MR, Rasolonjatovo IM, Reed MT, Rigatti R, Rodighiero C, Ross MT, Sabot A, Sankar SV, Scally A, Schroth GP, Smith ME, Smith VP, Spiridou A, Torrance PE, Tzonev SS, Vermaas EH, Walter K, Wu X, Zhang L, Alam MD, Anastasi C, Aniebo IC, Bailey DM, Bancarz IR, Banerjee S, Barbour SG, Baybayan PA, Benoit VA, Benson KF, Bevis C, Black PJ, Boodhun A, Brennan JS, Bridgham JA, Brown RC, Brown AA, Buermann DH, Bundu AA, Burrows JC, Carter NP, Castillo N, Chiara ECM, Chang S, Neil Cooley R, Crake NR, Dada OO, Diakoumakos KD, Dominguez-Fernandez B, Earnshaw DJ, Egbujor UC, Elmore DW, Etchin SS, Ewan MR, Fedurco M, Fraser LJ, Fuentes Fajardo KV, Scott Furey W, George D, Gietzen KJ, Goddard CP, Golda GS, Granieri PA, Green DE, Gustafson DL, Hansen NF, Harnish K, Haudenschild CD, Heyer NI, Hims MM, Ho JT, Horgan AM, Hoschler K, Hurwitz S, Ivanov DV, Johnson MQ, James T, Huw Jones TA, Kang GD, Kerelska TH, Kersey AD, Khrebtukova I, Kindwall AP, Kingsbury Z, Kokko-Gonzales PI, Kumar A, Laurent MA, Lawley CT, Lee SE, Lee X, Liao AK, Loch JA, Lok M, Luo S, Mammen RM, Martin JW, McCauley PG, McNitt P, Mehta P, Moon KW, Mullens JW, Newington T, Ning Z, Ling Ng B, Novo SM, O'Neill MJ, Osborne MA, Osnowski A, Ostadan O, Paraschos LL, Pickering L, Pike AC, Pike AC, Chris Pinkard D, Pliskin DP, Podhasky J, Quijano VJ, Raczy C, Rae VH, Rawlings SR, Chiva Rodriguez A, Roe PM, Rogers J, Rogert Bacigalupo MC, Romanov N, Romieu A, Roth RK, Rourke NJ, Ruediger ST, Rusman E, Sanches-Kuiper RM, Schenker MR, Seoane JM, Shaw RJ, Shiver MK, Short SW, Sizto NL, Sluis JP, Smith MA, Ernest Sohna Sohna J, Spence EJ, Stevens K, Sutton N, Szajkowski L,

Tregidgo CL, Turcatti G, Vandevondele S, Verhovsky Y, Virk SM, Wakelin S, Walcott GC, Wang J, Worsley GJ, Yan J, Yau L, Zuerlein M, Rogers J, Mullikin JC, Hurles ME, McCooke NJ, West JS, Oaks FL, Lundberg PL, Klenerman D, Durbin R, Smith AJ (2008) Accurate whole human genome sequencing using reversible terminator chemistry. Nature 456 (7218):53-59. doi:10.1038/nature07517

24. Fedurco M, Romieu A, Williams S, Lawrence I, Turcatti G (2006) BTA, a novel reagent for DNA attachment on glass and efficient generation of solid-phase amplified DNA colonies. Nucleic Acids Res 34 (3):e22. doi:10.1093/nar/gnj023

25. Bentley DR, Balasubramanian S, Swerdlow HP, Smith GP, Milton J, Brown CG, Hall KP, Evers DJ, Barnes CL, Bignell HR, Boutell JM, Bryant J, Carter RJ, Keira Cheetham R, Cox AJ, Ellis DJ, Flatbush MR, Gormley NA, Humphray SJ, Irving LJ, Karbelashvili MS, Kirk SM, Li H, Liu X, Maisinger KS, Murray LJ, Obradovic B, Ost T, Parkinson ML, Pratt MR, Rasolonjatovo IMJ, Reed MT, Rigatti R, Rodighiero C, Ross MT, Sabot A, Sankar SV, Scally A, Schroth GP, Smith ME, Smith VP, Spiridou A, Torrance PE, Tzonev SS, Vermaas EH, Walter K, Wu X, Zhang L, Alam MD, Anastasi C, Aniebo IC, Bailey DMD, Bancarz IR, Banerjee S, Barbour SG, Baybayan PA, Benoit VA, Benson KF, Bevis C, Black PJ, Boodhun A, Brennan JS, Bridgham JA, Brown RC, Brown AA, Buermann DH, Bundu AA, Burrows JC, Carter NP, Castillo N, Chiara E. Catenazzi M, Chang S, Neil Cooley R, Crake NR, Dada OO, Diakoumakos KD, Dominguez-Fernandez B, Earnshaw DJ, Egbujor UC, Elmore DW, Etchin SS, Ewan MR, Fedurco M, Fraser LJ, Fuentes Fajardo KV, Scott Furey W, George D, Gietzen KJ, Goddard CP, Golda GS, Granieri PA, Green DE, Gustafson DL, Hansen NF, Harnish K, Haudenschild CD, Heyer NI, Hims MM, Ho JT, Horgan AM, Hoschler K, Hurwitz S, Ivanov DV, Johnson MQ, James T, Huw Jones TA, Kang G-D, Kerelska TH, Kersey AD, Khrebtukova I, Kindwall AP, Kingsbury Z, Kokko-Gonzales PI, Kumar A, Laurent MA, Lawley CT, Lee SE, Lee X, Liao AK, Loch JA, Lok M, Luo S, Mammen RM, Martin JW, McCauley PG, McNitt P, Mehta P, Moon KW, Mullens JW, Newington T, Ning Z, Ling Ng B, Novo SM, O'Neill MJ, Osborne MA, Osnowski A, Ostadan O, Paraschos LL, Pickering L, Pike AC, Pike AC, Chris Pinkard D, Pliskin DP, Podhasky J, Quijano VJ, Raczy C, Rae VH, Rawlings SR, Chiva

Rodriguez A, Roe PM, Rogers J, Rogert Bacigalupo MC, Romanov N, Romieu A, Roth RK, Rourke NJ, Ruediger ST, Rusman E, Sanches-Kuiper RM, Schenker MR, Seoane JM, Shaw RJ, Shiver MK, Short SW, Sizto NL, Sluis JP, Smith MA, Ernest Sohna Sohna J, Spence EJ, Stevens K, Sutton N, Szajkowski L, Tregidgo CL, Turcatti G, vandeVondele S, Verhovsky Y, Virk SM, Wakelin S, Walcott GC, Wang J, Worsley GJ, Yan J, Yau L, Zuerlein M, Rogers J, Mullikin JC, Hurles ME, McCooke NJ, West JS, Oaks FL, Lundberg PL, Klenerman D, Durbin R, Smith AJ (2008) Accurate whole human genome sequencing using reversible terminator chemistry. Nature 456 (7218):53-59. doi:10.1038/nature07517

26. Guo J, Xu N, Li Z, Zhang S, Wu J, Kim DH, Sano Marma M, Meng Q, Cao H, Li X, Shi S, Yu L, Kalachikov S, Russo JJ, Turro NJ, Ju J (2008) Four-color DNA sequencing with *O*-modified nucleotide reversible terminators and chemically cleavable fluorescent dideoxynucleotides. Proceedings of the National Academy of Sciences 105 (27):9145-9150. doi:doi:10.1073/pnas.0804023105

27. Simpson JT, Workman RE, Zuzarte PC, David M, Dursi LJ, Timp W (2017) Detecting DNA cytosine methylation using nanopore sequencing. Nature Methods 14 (4):407-410. doi:10.1038/nmeth.4184

28. Levene MJ, Korlach J, Turner SW, Foquet M, Craighead HG, Webb WW (2003) Zero-mode waveguides for single-molecule analysis at high concentrations. Science 299 (5607):682-686. doi:10.1126/science.1079700

29. Haque F, Li J, Wu H-C, Liang X-J, Guo P (2013) Solid-state and biological nanopore for real-time sensing of single chemical and sequencing of DNA. Nano Today 8 (1):56-74. doi:https://doi.org/10.1016/j.nantod.2012.12.008

30. Wang Y, Yang Q, Wang Z (2015) The evolution of nanopore sequencing. Frontiers in Genetics 5. doi:10.3389/fgene.2014.00449

31. Churko JM, Mantalas GL, Snyder MP, Wu JC (2013) Overview of high throughput sequencing technologies to elucidate molecular pathways in cardiovascular diseases. Circulation Research 112 (12):1613-1623. doi:doi:10.1161/CIRCRESAHA.113.300939

32. Gautam SS, Kc R, Leong KW, Mac Aogáin M, O'Toole RF (2019) A step-by-step beginner's protocol for whole genome sequencing of human bacterial pathogens. J Biol Methods 6 (1):e110. doi:10.14440/jbm.2019.276

33. Mahajan MC, McLellan AS (2020) Whole-exome sequencing (WES) for Illumina short read sequencers using solution-based capture. Methods Mol Biol 2076:85-108. doi:10.1007/978-1-4939-9882-1_5

34. Dilliott AA, Farhan SMK, Ghani M, Sato C, Liang E, Zhang M, McIntyre AD, Cao H, Racacho L, Robinson JF, Strong MJ, Masellis M, Bulman DE, Rogaeva E, Lang A, Tartaglia C, Finger E, Zinman L, Turnbull J, Freedman M, Swartz R, Black SE, Hegele RA (2018) Targeted next-generation dequencing and bioinformatics pipeline to evaluate genetic determinants of constitutional disease. J Vis Exp (134). doi:10.3791/57266

35. Wheeler DA, Srinivasan M, Egholm M, Shen Y, Chen L, McGuire A, He W, Chen Y-J, Makhijani V, Roth GT, Gomes X, Tartaro K, Niazi F, Turcotte CL, Irzyk GP, Lupski JR, Chinault C, Song X-z, Liu Y, Yuan Y, Nazareth L, Qin X, Muzny DM, Margulies M, Weinstock GM, Gibbs RA, Rothberg JM (2008) The complete genome of an individual by massively parallel DNA sequencing. Nature 452 (7189):872-876. doi:10.1038/nature06884

36. Chou J, Ohsumi TK, Geha RS (2012) Use of whole exome and genome sequencing in the identification of genetic causes of primary immunodeficiencies. Curr Opin Allergy Clin Immunol 12 (6):623-628. doi:10.1097/ACI.0b013e3283588ca6

37. Bewicke-Copley F, Arjun Kumar E, Palladino G, Korfi K, Wang J (2019) Applications and analysis of targeted genomic sequencing in cancer studies. Comput Struct Biotechnol J 17:1348-1359. doi:10.1016/j.csbj.2019.10.004

38. Han S-W, Kim H-P, Shin J-Y, Jeong E-G, Lee W-C, Lee K-H, Won J-K, Kim T-Y, Oh D-Y, Im S-A, Bang Y-J, Jeong S-Y, Park KJ, Park J-G, Kang GH, Seo J-S, Kim J-I, Kim T-Y (2013) Targeted Sequencing of Cancer-Related Genes in Colorectal Cancer Using Next-Generation Sequencing. PLOS ONE 8 (5):e64271. doi:10.1371/journal.pone.0064271

39. Dongre HN, Haave H, Fromreide S, Erland FA, Moe SEE, Dhayalan SM, Riis RK, Sapkota D, Costea DE, Aarstad HJ, Vintermyr OK (2021) Targeted next-generation sequencing of cancer-related genes in a Norwegian patient cohort with head and neck squamous cell carcinoma reveals novel actionable mutations and correlations With pathological parameters. Frontiers in Oncology 11. doi:10.3389/fonc.2021.734134

40. Gulilat M, Lamb T, Teft WA, Wang J, Dron JS, Robinson JF, Tirona RG, Hegele RA, Kim RB, Schwarz UI (2019) Targeted next generation sequencing as a tool for precision medicine. BMC Medical Genomics 12 (1):81. doi:10.1186/s12920-019-0527-2

41. Lim B, Lin Y, Navin N (2020) Advancing cancer research and medicine with single-cell genomics. Cancer Cell 37 (4):456-470. doi:10.1016/j.ccell.2020.03.008

42. Evrony GD, Hinch AG, Luo C (2021) Applications of single-cell DNA sequencing. Annual Review of Genomics and Human Genetics 22 (1):171-197. doi:10.1146/annurev-genom-111320-090436

43. Hu P, Zhang W, Xin H, Deng G (2016) Single cell isolation and analysis. Frontiers in Cell and Developmental Biology 4. doi:10.3389/fcell.2016.00116

44. Evrony GD, Hinch AG, Luo C (2021) Applications of single-cell DNA sequencing. Annu Rev Genomics Hum Genet 22:171-197. doi:10.1146/annurev-genom-111320-090436

45. Lee J-H, Gao C, Peng G, Greer C, Ren S, Wang Y, Xiao X (2011) Analysis of transcriptome complexity through RNA sequencing in normal and failing murine hearts. Circulation Research 109 (12):1332-1341. doi:doi:10.1161/CIRCRESAHA.111.249433

46. Wang Z, Gerstein M, Snyder M (2009) RNA-Seq: a revolutionary tool for transcriptomics. Nature Reviews Genetics 10 (1):57-63. doi:10.1038/nrg2484

47. Edgren H, Murumagi A, Kangaspeska S, Nicorici D, Hongisto V, Kleivi K, Rye IH, Nyberg S, Wolf M, Borresen-Dale A-L, Kallioniemi O (2011) Identification of fusion genes in breast cancer by paired-end RNA-sequencing. Genome Biology 12 (1):R6. doi:10.1186/gb-2011-12-1-r6

48. Łabaj PP, Leparc GG, Linggi BE, Markillie LM, Wiley HS, Kreil DP (2011) Characterization and improvement of RNA-Seq precision in quantitative transcript expression profiling. Bioinformatics 27 (13):i383-391. doi:10.1093/bioinformatics/btr247

49. Sánchez-Pla A, Reverter F, Ruíz de Villa MC, Comabella M (2012) Transcriptomics: mRNA and alternative splicing. J Neuroimmunol 248 (1-2):23-31. doi:10.1016/j.jneuroim.2012.04.008

50. Hoeijmakers WA, Bártfai R, Stunnenberg HG (2013) Transcriptome analysis using RNA-Seq. Methods Mol Biol 923:221-239. doi:10.1007/978-1-62703-026-7_15

51. Rosenow C, Saxena RM, Durst M, Gingeras TR (2001) Prokaryotic RNA preparation methods useful for high density array analysis: comparison of two approaches. Nucleic Acids Res 29 (22):E112. doi:10.1093/nar/29.22.e112

52. Corchete LA, Rojas EA, Alonso-López D, De Las Rivas J, Gutiérrez NC, Burguillo FJ (2020) Systematic comparison and assessment of RNA-seq procedures for gene expression quantitative analysis. Scientific Reports 10 (1):19737. doi:10.1038/s41598-020-76881-x

53. Conesa A, Madrigal P, Tarazona S, Gomez-Cabrero D, Cervera A, McPherson A, Szcześniak MW, Gaffney DJ, Elo LL, Zhang X, Mortazavi A (2016) A survey of best practices for RNA-seq data analysis. Genome Biology 17 (1):13. doi:10.1186/s13059-016-0881-8

54. Hong M, Tao S, Zhang L, Diao L-T, Huang X, Huang S, Xie S-J, Xiao Z-D, Zhang H (2020) RNA sequencing: new technologies and applications in cancer research. Journal of Hematology & Oncology 13 (1):166. doi:10.1186/s13045-020-01005-x

55. Oshlack A, Robinson MD, Young MD (2010) From RNA-seq reads to differential expression results. Genome Biol 11 (12):220. doi:10.1186/gb-2010-11-12-220

56. Govindarajan M, Wohlmuth C, Waas M, Bernardini MQ, Kislinger T (2020) High-throughput approaches for precision medicine in high-grade serous ovarian cancer. J Hematol Oncol 13 (1):134. doi:10.1186/s13045-020-00971-6

57. Wu H, Li X, Li H (2019) Gene fusions and chimeric RNAs, and their implications in cancer. Genes Dis 6 (4):385-390. doi:10.1016/j.gendis.2019.08.002

58. Wang N, Zheng J, Chen Z, Liu Y, Dura B, Kwak M, Xavier-Ferrucio J, Lu Y-C, Zhang M, Roden C, Cheng J, Krause DS, Ding Y, Fan R, Lu J (2019) Single-cell microRNA-mRNA co-sequencing reveals non-genetic heterogeneity and mechanisms of microRNA regulation. Nature Communications 10 (1):95. doi:10.1038/s41467-018-07981-6

59. Pritchard CC, Cheng HH, Tewari M (2012) MicroRNA profiling: approaches and considerations. Nature Reviews Genetics 13 (5):358-369. doi:10.1038/nrg3198

60. Mannarapu M, Dariya B, Bandapalli OR (2021) Application of single-cell sequencing technologies in pancreatic cancer. Mol Cell Biochem 476 (6):2429-2437. doi:10.1007/s11010-021-04095-4

61. Li L, Xiong F, Wang Y, Zhang S, Gong Z, Li X, He Y, Shi L, Wang F, Liao Q, Xiang B, Zhou M, Li X, Li Y, Li G, Zeng Z, Xiong W, Guo C (2021) What are the applications of single-cell RNA sequencing in cancer research: a systematic review. Journal of Experimental & Clinical Cancer Research 40 (1):163. doi:10.1186/s13046-021-01955-1

62. Lamond AI, Earnshaw WC (1998) Structure and function in the nucleus. Science 280 (5363):547-553. doi:10.1126/science.280.5363.547

63. Buenrostro JD, Giresi PG, Zaba LC, Chang HY, Greenleaf WJ (2013) Transposition of native chromatin for fast and sensitive epigenomic profiling of open chromatin, DNA-binding proteins and nucleosome position. Nature Methods 10 (12):1213-1218. doi:10.1038/nmeth.2688

64. Breiling A, Lyko F (2015) Epigenetic regulatory functions of DNA modifications: 5-methylcytosine and beyond. Epigenetics & Chromatin 8 (1):24. doi:10.1186/s13072-015-0016-6

65. Kawakatsu T (2020) Whole-genome bisulfite sequencing and epigenetic variation in cereal methylomes. Methods Mol Biol 2072:119-128. doi:10.1007/978-1-4939-9865-4_10

66. Lister R, Pelizzola M, Kida YS, Hawkins RD, Nery JR, Hon G, Antosiewicz-Bourget J, O'Malley R, Castanon R, Klugman S, Downes M, Yu R, Stewart R, Ren B, Thomson JA, Evans RM, Ecker JR (2011) Hotspots of aberrant epigenomic reprogramming in human induced pluripotent stem cells. Nature 471 (7336):68-73. doi:10.1038/nature09798

67. Yan JG, Fu HY, Shen JZ, Zhou HR, Zhang YY, Huang JL, Chen CJ, Huang SH (2016) Application of bisulfite sequencing PCR in detecting the abnormal methylation of suppressor gene of Wnt signaling pathway in acute promyelocytic leukemia. Zhongguo Shi Yan Xue Ye Xue Za Zhi 24 (5):1299-1304. doi:10.7534/j.issn.1009-2137.2016.05.003

68. Xu H, Zhao Y, Liu Z, Zhu W, Zhou Y, Zhao Z (2012) Bisulfite genomic sequencing of DNA from dried blood spot microvolume samples. Forensic Sci Int Genet 6 (3):306-309. doi:10.1016/j.fsigen.2011.06.007

69. Eder T, Grebien F (2022) Comprehensive assessment of differential ChIP-seq tools guides optimal algorithm selection. Genome Biology 23 (1):119. doi:10.1186/s13059-022-02686-y

70. Mundade R, Ozer HG, Wei H, Prabhu L, Lu T (2014) Role of ChIP-seq in the discovery of transcription factor binding sites, differential gene regulation mechanism, epigenetic marks and beyond. Cell Cycle 13 (18):2847-2852. doi:10.4161/15384101.2014.949201

71. Nakato R, Sakata T (2021) Methods for ChIP-seq analysis: A practical workflow and advanced applications. Methods 187:44-53. doi:https://doi.org/10.1016/j.ymeth.2020.03.005

72. Dirks RA, Stunnenberg HG, Marks H (2016) Genome-wide epigenomic profiling for biomarker discovery. Clin Epigenetics 8:122. doi:10.1186/s13148-016-0284-4

73. Chen Y, Zhang Y, Wang Y, Zhang L, Brinkman EK, Adam SA, Goldman R, van Steensel B, Ma J, Belmont AS (2018) Mapping 3D genome organization relative to nuclear compartments using TSA-Seq as a cytological ruler. Journal of Cell Biology 217 (11):4025-4048. doi:10.1083/jcb.201807108

74. Tanizawa H, Noma K (2012) Unravelling global genome organization by 3C-seq. Semin Cell Dev Biol 23 (2):213-221. doi:10.1016/j.semcdb.2011.11.003

75. Dekker J, Rippe K, Dekker M, Kleckner N (2002) Capturing chromosome conformation. Science 295 (5558):1306-1311. doi:doi:10.1126/science.1067799

76. Rebouissou C, Sallis S, Forné T (2022) Quantitative chromosome conformation capture (3C-qPCR). Methods Mol Biol 2532:3-13. doi:10.1007/978-1-0716-2497-5_1

77. van Opijnen T, Bodi KL, Camilli A (2009) Tn-seq: high-throughput parallel sequencing for fitness and genetic interaction studies in microorganisms. Nat Methods 6 (10):767-772. doi:10.1038/nmeth.1377

78. Carabetta VJ, Esquilin-Lebron K, Zelzion E, Boyd JM (2021) Genetic approaches to uncover gene products involved in iron-sulfur protein maturation: high-throughput genomic screening using transposon sequencing. In: Dos Santos PC (ed) Fe-S Proteins: Methods and Protocols. Springer US, New York, NY, pp 51-68. doi:10.1007/978-1-0716-1605-5_3

79. Barquist L, Mayho M, Cummins C, Cain AK, Boinett CJ, Page AJ, Langridge GC, Quail MA, Keane JA, Parkhill J (2016) The TraDIS toolkit: sequencing and analysis for dense transposon mutant libraries. Bioinformatics 32 (7):1109-1111. doi:10.1093/bioinformatics/btw022

80. Cain AK, Barquist L, Goodman AL, Paulsen IT, Parkhill J, van Opijnen T (2020) A decade of advances in transposon-insertion sequencing. Nature Reviews Genetics 21 (9):526-540. doi:10.1038/s41576-020-0244-x

81. Thibault D, Jensen PA, Wood S, Qabar C, Clark S, Shainheit MG, Isberg RR, van Opijnen T (2019) Droplet Tn-Seq combines microfluidics with Tn-Seq for identifying complex single-cell phenotypes. Nat Commun 10 (1):5729. doi:10.1038/s41467-019-13719-9

82. Yasir M, Turner AK, Bastkowski S, Baker D, Page AJ, Telatin A, Phan MD, Monahan L, Savva GM, Darling A, Webber MA, Charles IG (2020) TraDIS-Xpress: a high-resolution whole-genome assay identifies novel mechanisms of triclosan action and resistance. Genome Res 30 (2):239-249. doi:10.1101/gr.254391.119

83. Lam HYK, Clark MJ, Chen R, Chen R, Natsoulis G, O'Huallachain M, Dewey FE, Habegger L, Ashley EA, Gerstein MB, Butte AJ, Ji HP, Snyder M (2012) Performance comparison of whole-genome sequencing platforms. Nature Biotechnology 30 (1):78-82. doi:10.1038/nbt.2065

84. Ivády G, Madar L, Dzsudzsák E, Koczok K, Kappelmayer J, Krulisova V, Macek M, Jr., Horváth A, Balogh I (2018) Analytical parameters and validation of homopolymer detection in a pyrosequencing-based next generation sequencing system. BMC Genomics 19 (1):158. doi:10.1186/s12864-018-4544-x

85. Zook JM, Salit M (2011) Genomes in a bottle: creating standard reference materials for genomic variation - why, what and how? Genome Biology 12 (1):P31. doi:10.1186/gb-2011-12-s1-p31

86. Brownstein CA, Beggs AH, Homer N, Merriman B, Yu TW, Flannery KC, DeChene ET, Towne MC, Savage SK, Price EN, Holm IA, Luquette LJ, Lyon E, Majzoub J, Neupert P, McCallie D, Jr., Szolovits P, Willard HF, Mendelsohn NJ, Temme R, Finkel RS, Yum SW, Medne L, Sunyaev SR, Adzhubey I, Cassa CA, de Bakker PI, Duzkale H, Dworzyński P, Fairbrother W, Francioli L, Funke BH, Giovanni MA, Handsaker RE, Lage K, Lebo MS, Lek M, Leshchiner I, MacArthur DG, McLaughlin HM, Murray MF, Pers TH, Polak PP, Raychaudhuri S, Rehm HL, Soemedi R, Stitziel NO, Vestecka S, Supper J, Gugenmus C, Klocke B, Hahn A, Schubach M, Menzel M, Biskup S, Freisinger P, Deng M, Braun M, Perner S, Smith RJ, Andorf JL, Huang J, Ryckman K, Sheffield VC, Stone EM, Bair T, Black-Ziegelbein EA, Braun TA, Darbro B, DeLuca AP, Kolbe DL, Scheetz TE, Shearer AE, Sompallae R, Wang K, Bassuk AG, Edens E, Mathews K, Moore SA, Shchelochkov OA, Trapane P, Bossler A, Campbell CA, Heusel JW, Kwitek A, Maga T, Panzer K, Wassink T, Van Daele D, Azaiez H, Booth K, Meyer N, Segal MM, Williams MS, Tromp G, White P, Corsmeier D, Fitzgerald-Butt S, Herman G, Lamb-Thrush D, McBride KL, Newsom D, Pierson CR, Rakowsky AT, Maver A, Lovrečić L, Palandačić A, Peterlin B, Torkamani A, Wedell A, Huss M, Alexeyenko A, Lindvall JM, Magnusson M, Nilsson D, Stranneheim H, Taylan F, Gilissen C, Hoischen A, van Bon B, Yntema H, Nelen

M, Zhang W, Sager J, Zhang L, Blair K, Kural D, Cariaso M, Lennon GG, Javed A, Agrawal S, Ng PC, Sandhu KS, Krishna S, Veeramachaneni V, Isakov O, Halperin E, Friedman E, Shomron N, Glusman G, Roach JC, Caballero J, Cox HC, Mauldin D, Ament SA, Rowen L, Richards DR, San Lucas FA, Gonzalez-Garay ML, Caskey CT, Bai Y, Huang Y, Fang F, Zhang Y, Wang Z, Barrera J, Garcia-Lobo JM, González-Lamuño D, Llorca J, Rodriguez MC, Varela I, Reese MG, De La Vega FM, Kiruluta E, Cargill M, Hart RK, Sorenson JM, Lyon GJ, Stevenson DA, Bray BE, Moore BM, Eilbeck K, Yandell M, Zhao H, Hou L, Chen X, Yan X, Chen M, Li C, Yang C, Gunel M, Li P, Kong Y, Alexander AC, Albertyn ZI, Boycott KM, Bulman DE, Gordon PM, Innes AM, Knoppers BM, Majewski J, Marshall CR, Parboosingh JS, Sawyer SL, Samuels ME, Schwartzentruber J, Kohane IS, Margulies DM (2014) An international effort towards developing standards for best practices in analysis, interpretation and reporting of clinical genome sequencing results in the CLARITY Challenge. Genome Biol 15 (3):R53. doi:10.1186/gb-2014-15-3-r53

87. Shendure J, Balasubramanian S, Church GM, Gilbert W, Rogers J, Schloss JA, Waterston RH (2017) DNA sequencing at 40: past, present and future. Nature 550 (7676):345-353. doi:10.1038/nature24286

88. Prober JM, Trainor GL, Dam RJ, Hobbs FW, Robertson CW, Zagursky RJ, Cocuzza AJ, Jensen MA, Baumeister K (1987) A system for rapid DNA sequencing with fluorescent chain-terminating dideoxynucleotides. Science 238 (4825):336-341. doi:10.1126/science.2443975

89. Craxton M (1991) Linear amplification sequencing, a powerful method for sequencing DNA. Methods 3 (1):20-26. doi:https://doi.org/10.1016/S1046-2023(05)80159-8

90. DeAngelis MM, Wang DG, Hawkins TL (1995) Solid-phase reversible immobilization for the isolation of PCR products. Nucleic Acids Res 23 (22):4742-4743. doi:10.1093/nar/23.22.4742

91. Zhang J, Fang Y, Hou JY, Ren HJ, Jiang R, Roos P, Dovichi NJ (1995) Use of non-cross-linked polyacrylamide for four-color DNA sequencing by capillary electrophoresis separation of fragments up to 640 bases in length in two hours. Anal Chem 67 (24):4589-4593. doi:10.1021/ac00120a026

92. Pereira R, Oliveira J, Sousa M (2020) Bioinformatics and computational tools for next-generation sequencing analysis in clinical genetics. J Clin Med 9 (1). doi:10.3390/jcm9010132

93. Kanzi AM, San JE, Chimukangara B, Wilkinson E, Fish M, Ramsuran V, de Oliveira T (2020) Next generation sequencing and bioinformatics analysis of family genetic inheritance. Front Genet 11:544162. doi:10.3389/fgene.2020.544162

**Tables**

**Table 1. Comparing the properties of different NGS technologies**

| Platform | Mechanism/Chemistry | Year | Read length (bp) | Output/day (Mb) | Common error type | Cost/Mb |
|---|---|---|---|---|---|---|
| Automated Sanger | Reversible dye terminator | 2002 | 800 | 6 | NA | $500.00 |
| 454 (Roche) | EmPCR, PP$_i$ release, Luminescence | 2005 | 100-500 | 700 | InDel | $20.00 |
| Ion Torrent (Life Technologies) | EmPCR, Proton release | 2011 | 100-400 | 8000 | InDel | $0.50 |
| Illumina | Bridge amplification PCR, dye terminator | 2011 | 100-300 | 50,000 | Substitution | $0.50 |
| SOLiD (Applied Biosystems) | PCR, Octomer ligation | 2011 | 25-75 | 5000 | Mismatch | $0.50 |
| SMRT (Pacific Biosciences) | Zero-mode wavelength, SMS | 2011 | 14,000-60,000 | 1000 | InDel/mismatch | $1.00 |
| Nanopore (Oxford) | Membrane nanopores, SMS | 2015 | 150000 | > 90 | InDel/mismatch | $2.00 |

Abbreviations: bp: base pairs; Mb: megabases; EmPCR: emulsion PCR, PP$_i$: pyrophosphate, SMS: single molecule sequencing; InDel: insertion-deletion mutation. Data used in this table is from references 10, 11, 16, and 31.

**Table 2. Notable improvements in NGS technology**

| Improvement [87] | Consequence |
|---|---|
| Use of dye labeled terminators instead of labeled primers | Allowed for one sequencing reaction per sample instead of multiple reactions. |
| Use of mutant form of T7 DNA polymerase | Effectviely Incorporated dye-labelled terminators. |
| Use of PCR amplification | Reduced required template concentration and aided miniaturization. |
| Use of magnetic beads for DNA purification | Simplified the automation of pre-sequencing steps. |
| The abiltity to sequence double-stranded DNA | Enabled the use of plasmid clones and allowed for paired end sequencing. |
| Use of capillary electrophoresis | Eliminated use of agarose gels, and simplified the extraction and interpretation of the fluorescent signals. |
| Comercialization of platforms | Improved efficiencies and reduced error rates. |

**Figures**

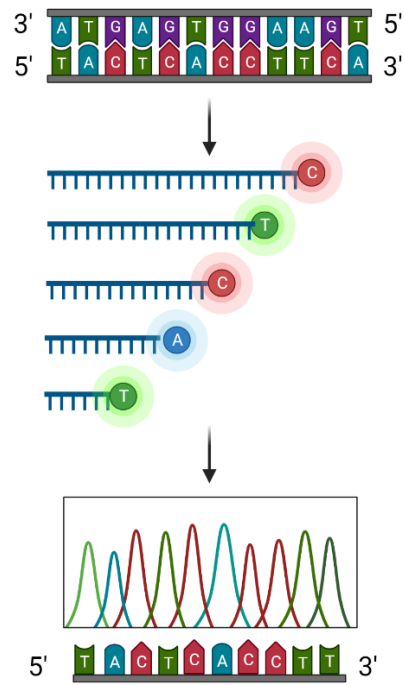

**Figure 1.** Sanger sequencing workflow. The reverse strand of linear DNA is synthesized using DNA polymerase and four fluorescently labeled dideoxyribonucleotides (ddNTPs), which when incorporated will terminate chain elongation and emit a fluorescence signal unique to each ddNTP. The resulting fragments are analyzed via capillary gel electrophoresis. Created with BioRender.com.

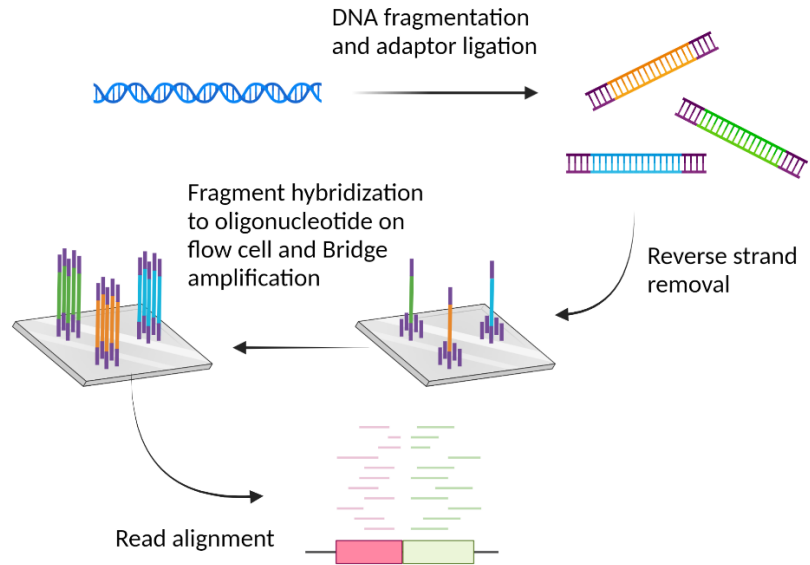

**Figure 2**. 454/Roche and Ion Torrent sequencing. DNA is fragmented and attached to a bead via ligated adapters sequences. Amplification of the fragment attached to the beads occurs by emulsion PCR (emPCR). Beads (clones) are collected in wells of picotiter plates followed by the addition of DNA polymerase and dNTPs. For 454/Roche sequencing, the sequence is determined by measuring the light signal, which is unique to each dNTP, emitted using a charged couple device (CCD) camera, following pyrophosphate ($PP_i$) release during base incorporation and extension. Ion Torrent sequencing follows a similar workflow, but sequence determination is made by measuring $H^+$ ion release during incorporation by a pH sensor. Created with BioRender.com.

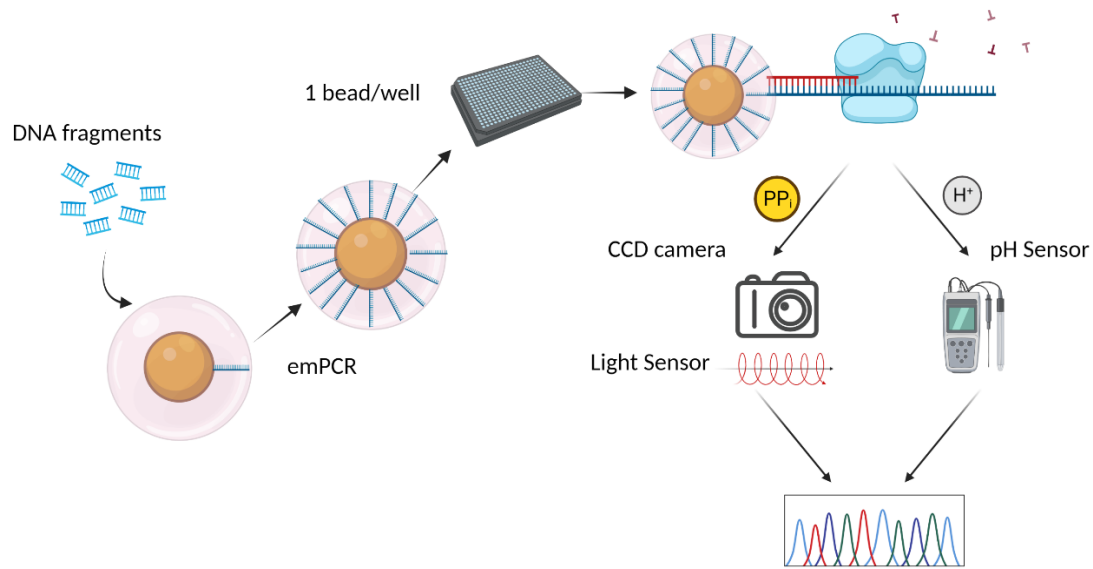

**Figure 3**. Illumina sequencing platform. DNA is fragmented, linearized, and ligated to adapters that will bind surface-tethered oligos via complementarity on a flow cell. The DNA fragments bend over to form bridges, followed by bridge amplification and cluster formation. Sequences are determined following incorporation of reversible dye ddNTPs. The fragments in the library are then assembled using various bioinformatic pipelines. Created with BioRender.com.

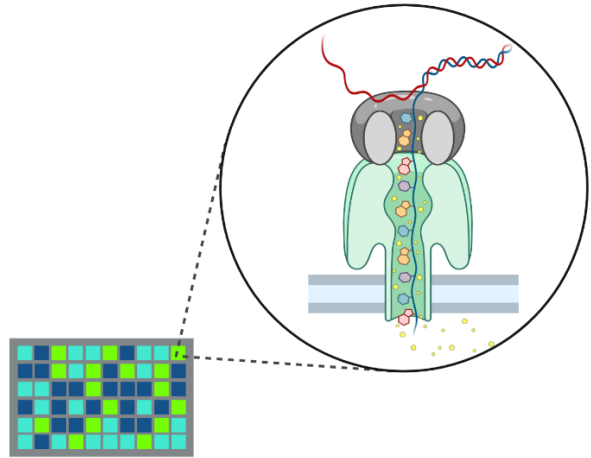

**Figure 4**. Oxford Nanopore sequencing. A ssDNA fragment is pushed through a membrane nanopore composed of a motor enzyme. Each nucleotide will alter the current through the pore by a different magnitude as it moves through the pore, allowing for sequence determination. Created with BioRender.com.